\documentclass{ppai}

\usepackage{ppai}

\usepackage[utf8]{inputenc} 
\usepackage[T1]{fontenc}    
\usepackage{hyperref}       
\usepackage{url}            
\usepackage{booktabs}       
\usepackage{amsfonts}       
\usepackage{nicefrac}       
\usepackage{microtype}      
\usepackage{lipsum}
\usepackage{graphicx}
\graphicspath{ {./images/} }
\usepackage{booktabs}
\usepackage{longtable}
\usepackage{color}
\usepackage{comment}
\usepackage{multicol}
\usepackage{enumitem}
\usepackage{wrapfig} 

\usepackage{listings}

\title{Audio Editing Features as User-Centric Privacy Defenses Against Large Language Model (LLM)-Based Emotion Inference Attacks}

\author{%
  Mohd. Farhan Israk Soumik \\
  School of Computing\\
  Southern Illinois University\\
  Carbondale, IL, 62901 \\
  \texttt{mohdfarhanisrak.soumik@siu.edu} \\
   \And
  W.K.M Mithsara \\
  School of Computing\\
  Southern Illinois University\\
  Carbondale, IL, 62901 \\
  \texttt{malithi.mithsara@siu.edu} \\
   \And
   Abdur R. Shahid \\
   School of Computing\\
  Southern Illinois University\\
  Carbondale, IL, 62901 \\
   \texttt{shahid@cs.siu.edu} \\
   \AND
  Ahmed Imteaj\\
  School of Computing\\
  Southern Illinois University\\
  Carbondale, IL, 62901 \\
   \texttt{imteaj@cs.siu.edu} \\
}

\begin{document}

\maketitle

\begin{abstract}
The rapid proliferation of speech-enabled technologies, including virtual assistants, video conferencing platforms, and wearable devices, has raised significant privacy concerns, particularly regarding the inference of sensitive emotional information from audio data. Existing privacy-preserving methods often compromise usability and security, limiting their adoption in practical scenarios. This paper introduces a novel, user-centric approach that leverages familiar audio editing techniques, specifically pitch and tempo manipulation, to protect emotional privacy without sacrificing usability. By analyzing popular audio editing applications on Android and iOS platforms, we identified these features as both widely available and usable. We rigorously evaluated their effectiveness against a threat model, considering adversarial attacks from diverse sources, including Deep Neural Networks (DNNs), Large Language Models (LLMs), and and reversibility testing. Our experiments, conducted on three distinct datasets, demonstrate that pitch and tempo manipulation effectively obfuscates emotional data. Additionally, we explore the design principles for lightweight, on-device implementation to ensure broad applicability across various devices and platforms.

\end{abstract}
\section{Introduction}

The interconnection between human emotion and audio is profound and intricate. Audio can evoke, amplify, and manipulate feelings through tones, rhythms, and frequencies~\cite{li2019music}. Acoustic details and voice inflections convey emotions more effectively than words. The prevalence of audio in our daily lives has increased due to advancements in modern communication technologies, particularly high-speed internet and mobile networking like 4G and 5G. These technologies have enabled the transmission of superior audio quality and seamless transmission, which has become the cornerstone of platforms like video conferencing, podcasts, and Voice-over-IP(VoIP) services. Video conferencing and streaming platforms like Zoom, Google Meet, Twitch, and social media have contributed to this expansion. AI-powered Personalized Virtual Assistants(e.g Alexa, Siri) and Large Language Models (LLMs) are gaining ground, making voice commands and interactions more ubiquitous while increasing the significance of audio as a data format\cite{hoy2018alexa}. We are slowly embracing the world of virtual reality(VR), augmented reality(AR), and immersive technologies, which are considered Extended Reality~\cite{begault20003}\cite{vr1}. Audio is a critical component for creating fully immersive experiences. Spatial audio, which lets users perceive sound directionally in 3D, has become crucial in AR and VR environments for more realistic and engaging interactions. In VR gaming, dynamic audio effects synchronize sound with virtual objects and movements, enhancing immersion. Machine learning optimizes audio processing for immersive technologies, including real-time noise cancellation, voice recognition, and personalized soundscapes.

As new disruptive technologies proliferate, audio gains relevance but also brings emerging security concerns. Audio recordings reveal content and sensitive private information, like emotions or geographic locations, from background noise and vocal cues, emphasizing audio privacy \cite{williams2023new}. Systems like Siri and Alexa use Speech Emotion Recognition (SER) to infer users’ affective states for applications like mental health monitoring and interactive entertainment\cite{alsenani2023privacy}. However, these systems are vulnerable to inference attacks that extract sensitive attributes like identity and gender from audio signals\cite{alsenani2023privacy}. Even Federated Learning, considered more secure, has been shown to be susceptible to inference attacks on audio data\cite{feng2021attribute}. Video conferencing platforms are also susceptible to attacks that extract personal attributes from audio, video, drawing, and writing data\cite{kagan, shahiddp}. Emerging VR devices present additional challenges due to their extensive sensor integrations, including audio. Attackers can exploit vocalizations in VR environments to fingerprint users and infer sensitive attributes like age, emotion, and personality using machine learning\cite{Munilla_Garrido_2024, alkaeed2023privacypreservationartificialintelligence, sandeepa}.

To mitigate the inference of sensitive attributes and information from speech data, researchers have proposed diverse privacy-preserving methods. These include sanitizing sensitive features to obscure private information, employing cryptographic techniques, and incorporating privacy-by-design principles to balance privacy and utility. Techniques such as differential privacy, disentangled representation learning, and deep neural network-based obfuscation have been explored to protect against attribute inference attacks, including gender and emotion recognition \cite{10740800, 10.1145/3411495.3421355, 9925054}. Frameworks leveraging advanced methods like hierarchical noise addition, task-relevant data transformation, and genetic programming have also been proposed to ensure privacy in speech-based systems, including federated learning, voice user interfaces, and smart speakers \cite{10.1145/3610887}. If we closely analyze these works, it becomes evident that they primarily achieve privacy by identifying and separating sensitive features of speech audio data using machine learning models. These models often involve computationally intensive architectures, such as autoencoders, to extract and isolate sensitive attributes. Once identified, these attributes are privatized using security measures like encryption or differential privacy. While some approaches allow users to define varying levels of privacy (e.g., high, medium, low) for different types of sensitive attributes, this often requires an understanding of the technical implications of these settings. Such reliance on user expertise or awareness may limit the accessibility and usability of these solutions, especially for non-technical users.

This paper proposes a novel approach to privacy preservation in audio-based emotion recognition attacks. Unlike existing works, it focuses on usability and aims to develop a method that can be adopted by users with minimal training or technical expertise. WE envisioned that the approach will be system-wide and not confined to specific applications or platforms, such as Apple Intelligence\cite{apple_intelligence}. Equally important, utilizing such features to modify audio files will have minimal impact on user acceptability, as users are already employing these features for fun and utility, thereby blurring the gap between usability and privacy protection.


To achieve this goal, we begin our study by analyzing audio editing applications available on the Google Play Store\cite{google_play} and Apple App Store\cite{apple_app_store} to identify common audio editing features. The rationale for this step lies in the fact that billions of people already use devices integrated into either the Android\cite{android} or iOS\cite{ios_18} ecosystems. We hypothesize that leveraging familiar tools and workflows will enhance the usability and adoption of our proposed approach. Our analysis reveals that many of these applications include features to modify pitch and tempo values through intuitive means, such as sliders. Building on this insight, we explore the potential of using systematic pitch and tempo manipulation as the foundation of a user-centric privacy-preservation mechanism. To rigorously evaluate the feasibility of this approach, we design a threat model that incorporates a diverse range of attack vectors, including various AI models based on Deep Neural Networks (DNNs), Large Language Models (LLMs), and reversibility testing using audio editing applications. In the evaluation, we utilize three distinct datasets to assess the generalizability of the proposed features in protecting privacy. Furthermore, we investigate potential biases (e.g., demographic biases) in the effectiveness of privacy protection and evaluate the approach's resilience against various attack types. Our extensive experiments demonstrate promising results, indicating that pitch and tempo manipulation can serve as an effective, user-friendly privacy-preserving mechanism against emotion recognition attacks. These findings pave the way for the development of practical, usable privacy-preserving defenses to safeguard sensitive data from audio files. In a nutshell, the main contributions of our paper are as follows. 

\textbf{User-Centric Privacy Against Emotion Inference Attacks.} This paper introduces a novel, user-centric approach to privacy preservation against audio-based emotion recognition attacks with focus on user's usability through existing tools and concepts.

\textbf{Leveraging Familiar Audio Features as Basis for Privacy-Preservation.} The proposed approach leverages intuitive audio-editing features such as pitch and tempo manipulation, widely available in existing ecosystems.

\textbf{Extensive Threat Model for Emotion Inference Attacks.} A rigorous threat model is designed, encompassing diverse attack vectors, including Deep Neural Networks (DNNs), Large Language Models (LLMs), and reversibility testing, to rigorously evaluate the proposed approach.

\textbf{Experimental Validation and Generalizability}. Extensive experiment conducted by carrying out pitch and tempo manipulation directly via a popular app on three different datasets demonstrating the effectiveness, generaliazibility, and robustness of the proposed method.

\section{Related Work}
\textbf{Emotion Inference Attacks on Audio Data:} Emotion inference attacks significantly threaten the privacy and security of systems that process emotional data. These attacks exploit machine learning models to infer sensitive, emotional states or related attributes without user consent, raising concerns in applications like virtual assistants, healthcare, and surveillance. Understanding their vulnerabilities to such attacks is critical as emotion recognition technologies become increasingly integrated into everyday systems. Most previous studies on attribute inference attacks in speech emotion recognition focus on scenarios where adversaries exploit model outputs to deduce private speaker attributes, such as identity, gender, or demographic traits, beyond the intended emotion classification\cite{10740800, feng2022user, zhao2023privacy}. Our study focuses on changing the tempo and pitch in audio by using uptempo based on usable privacy and eventually changing the emotion of the video. Judith et al. \cite{ley2022effects} propose a method that involves altering pitch in simple tones across different frequency ranges to measure people's emotional states by recording their body movements. Alica et al. \cite{fernandez2016influence} investigate the impact of tempo (measured in beats per minute) and rhythmic units (patterns like whole, half, eighth, and sixteenth notes) on listeners' emotional perceptions and mood regulation. It aims to understand how these factors induce positive or negative emotional states. Zhiyuan et al\cite{yu2023smack} introduces a novel class of adversarial attacks on voice-controlled systems (VCS), named SMACK, which manipulates semantic speech attributes (like prosody) to evade state-of-the-art defenses by combining genetic algorithms with gradient estimation for prosody optimization.

\textbf{Emotion Recognition Through Conventional Machine Learning Approaches and  Large Language Models on Audio Data:}s Mohd et al.\cite{10.1145/3652037.3663943} explore emotion recognition through audio and textual modalities using a Bi-LSTM model for audio data and a fine-tuned DistilRoBERTa model for textual data and they utilize the SAVEE, RAVDESS, and TESS datasets for analysis. Theresa et al. \cite{10.1145/3615834.3615837} propose a novel approach, MATS2L, to improve emotion recognition using physiological signals through self-supervised learning (SSL) and 1DCNN and using the WESAD dataset. Mandeep et al.\cite{10351697, singh2020emotion} explore emotion recognition by processing audio spectrograms and video frames using deep learning architectures, including a 3D CNN for video frames and a CNN+RNN for audio spectrograms. Their work utilizes the IEMOCAP dataset, which comprises 12 hours of audiovisual data from 10 actors expressing various emotions, such as anger, happiness, sadness, and neutrality. When considering conventional methods, machine learning techniques have reached a significant milestone in accurately identifying emotions, demonstrating remarkable advancements in the field\cite{10351697, singh2020emotion}. However, regarding utilizing large language models (LLMs) for emotion recognition, only a few studies have been conducted, leaving this area relatively under-explored \cite{wu2024beyond, xu2024secap}. Zehui et al.\cite{wu2024beyond} explore a novel approach to integrating speech characteristics into Large Language Models (LLMs) for emotion recognition. The study focuses on enhancing the multi-modal capabilities of LLMs without requiring architectural modifications by converting audio features such as volume, pitch, and speaking rate into natural language descriptions. The researchers evaluate various LLMs (e.g., LLaMA-2, LLaMA-3, Phi-3) using LoRA fine-tuning and IEMOCAP and MELD datasets. Yaoxun et al. \cite{xu2024secap} introduce a novel framework, SECap, for describing speech emotions using natural language, moving beyond traditional classification into predefined emotion categories. They utilize HuBERT as the audio encoder to extract robust speech features, while the Bridge-Net is employed to process these features, separating emotion-related acoustic information from content. LLaMA is then used to generate coherent and human-like emotion captions. The study utilizes the EMOSpeech dataset, which consists of 41.6 hours of annotated speech data, including corresponding emotion captions, transcriptions, and labels. 

\textbf{Privacy-Preserving Mechanisms to Defend Against Emotion Inference Attacks:} Privacy-preserving mechanisms aim to mitigate these threats by sanitizing sensitive features, applying cryptographic techniques, or incorporating privacy-by-design principles. These methods ensure that emotion recognition systems retain functionality while protecting users’ emotional data from unauthorized inference, balancing utility and privacy in sensitive applications. Haijiao et al.\cite{10740800} address privacy challenges in federated learning (FL) for Speech Emotion Recognition by introducing a Gradient-level Hierarchical Differential Privacy (GHDP) framework by employing gradient normalization, clipping significant gradients, and adding hierarchical noise to early model layers during backpropagation to mitigate attribute inference attacks, such as gender prediction. Ranya et al.\cite{10.1145/3411495.3421355} explore privacy challenges in voice user interfaces (VUIs), which are widely used in smartphones, home assistants, and IoT devices. The focus is on defending against attribute inference attacks that exploit deep acoustic models to infer private attributes like gender or emotion. The authors propose a novel framework utilizing disentangled representation learning to filter sensitive information while retaining utility for primary tasks such as speech recognition.  Wang et al.\cite{9925054} proposes a novel privacy-preserving framework for audio-based applications by implementing a deep neural network (DNN)-based obfuscator to transform raw audio into a privacy-preserving format, retaining only task-relevant data without taking sensitive attributes like gender, emotion, or identity. This allows users to authorize only specific tasks while obfuscating unauthorized information. Brian et al\cite{10.1145/3610887} introduce DARE-GP (Defeating Acoustic Recognition of Emotion via Genetic Programming), a novel framework to protect user privacy from unauthorized speech emotion recognition (SER) systems, such as those embedded in smart speaker voice assistants (e.g., Amazon Echo, Google Home).

\textit{\textbf{Observations:} Current methods for defending against emotion inference attacks are heavily reliant on computationally intensive models, such as deep learning and disentangled representation learning, which focus on separating sensitive features like emotions. While effective, these approaches often overlook usability and adaptability, and hence limits their practical deployment for non-technical users. Furthermore, techniques like pitch and tempo manipulation, which are intuitive and accessible, remain underexplored for privacy preservation. Current frameworks also lack emphasis on cross-platform applicability and usability, particularly in diverse systems like virtual assistants and VR devices. Our work addresses these limitations by leveraging familiar audio-editing features to design lightweight, user-centric privacy mechanisms that balance effectiveness and accessibility.}


\section{Threat Model}
\textbf{Assumption on Target Systems:} We consider target systems that interact with speech data in various capacities. These include any system that records, stores, processes, or provides third-party access to speech data. Examples of such systems include virtual assistants, such as Alexa\cite{alexa}, Siri\cite{siri}, and Google Assistant\cite{google_assistant}, which process user input to perform tasks or provide services, and large language model-based chatbots, such as ChatGPT\cite{chatgpt} and Gemini\cite{gemini_google}, which may handle audio data as part of their multimodal interaction capabilities. Video conferencing platforms, such as Google Meet\cite{google_meet} and Zoom\cite{zoom}, that record or store audio for communication purposes are also included. Also, cloud-based audio applications, which store user data in cloud infrastructures, pose risks such as unauthorized access, data breaches, and malicious exploitation. Emerging technologies like Meta Smart Glasses\cite{meta_smart_glasses}, which integrate audio for real-time communication, transcription, or augmented reality, process and potentially store speech data, raising privacy concerns due to their pervasive nature and continuous data collection. 

\textbf{Adversaries and Attack Vectors:} The mentioned systems can be targeted by various adversaries with different motivations and methods. External hackers can breach cloud storage systems or intercept communication channels to extract speech data for unauthorized analysis or dissemination. Application developers also pose a potential risk when they have access to users' audio data. Developers may misuse this data for purposes beyond original intent for monetization through emotion recognition anslysis or sharing it with third parties without the explicit consent of users. Similarly, insider threats, such as colluding employees or individuals within an organization, with access to the sensitive data, may misuse their access for personal or financial gain or collaborate with external parties to exploit the data. 

\textbf{Assumption on Attacker's Knowledge:} We assume that the attackers posses significant knowledge and resources to compromise the privacy of speech data. This include,  knowledge of the application used to manipulate the speech data, including its features or method used; expertise and resources to develop or access advanced AI models specifically designed for inference attacks; and access to state-of-the-art LLMs like GPT-4o, which can analyze audio files to infer sensitive data.

\section{The Proposed Framework}
\subsection{Design Goals}
\begin{wrapfigure}{r}{0.5\textwidth} 
    \centering \vspace{-91pt}
    \includegraphics[width=\linewidth]{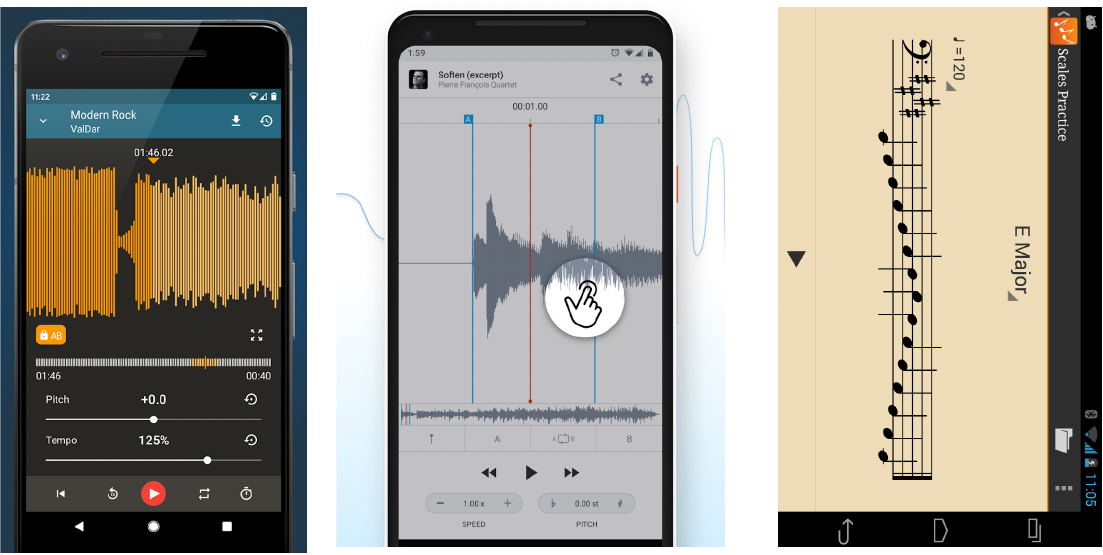} 
    \caption{Interface of audio editing apps: Up Tempo, AudioStretch, and Scales Practice. (Images taken from the apps' sites.)} \vspace{-15pt}
    \label{fig:enter-label}
\end{wrapfigure}
The core challenge in preserving privacy against the threat model lies in designing a mechanism that effectively protect sensitive emotional information while maintaining usability, adaptability, and compatibility across diverse systems and devices. This necessitates addressing several critical factors, including computational efficiency, user accessibility, and system-wide applicability. These goals include: \textit{(Lightweight On-Device Implementation)} The privacy-preserving mechanism should be computationally lightweight so that it can be implemented directly on user devices without requiring extensive processing power or memory. This is critical for devices with limited resources, such as smartphones, smart glasses, or IoT devices. \textit{(Usability)} The solution should prioritize ease of use which will allow users to adopt privacy-preserving features with minimal technical expertise or training. This involves intuitive interfaces and seamless integration with existing workflows to reduce barriers to adoption. \textit{(Adaptability)} It is highly desirable that the mechanism should be adaptable to a wide range of applications and scenarios, including virtual assistants, video conferencing platforms, and cloud-based services. Furthermore, it should accommodate varying data distribution and device capabilities for consistent performance across heterogeneous systems. Furthermore, it should be compatible with a wide range of devices and platforms, including Android, Windows, iOS, etc. \textit{(Robustness Against Attacks)} The design must be resilient to various attack vectors, including reversibility attempts, AI-based inference models, and adversaries leveraging large language models (LLMs).

\subsection{The Usable Privacy-Preserving Mechanism}

Our methodology to achieve the above mentioned design goals is rooted in usable security and privacy. Our objective is to develop privacy-preserving defense mechanisms based on features that users are already familiar with, which will enable intuitive adoption and minimizes the learning curve.
\begin{table}
    \centering \small
     \caption{Example popular audio editing apps and their key features, including pitch and tempo manipulation.}
    \begin{tabular}{c|c|c|c} 
    \hline
        App Name & Platform & Key Features & Popularity\\ \hline
       Up Tempo  & Android/iOS & Pitch adjustment, tempo control  & 1 Million+ downloads on Android, Rating 5 on iOS\\ \hline
       Scales Practice  & Android/iOS &  Pitch adjustment, tempo control & 100K+ Downloads on Android, 4.8 rating on iOS\\ \hline
       AudioStretch & Android & Pitch adjustment, speed control & 1 Million+ downloads with 3.5 average rating at playstore. \\ \hline
        BandLab & Android & Pitch and Tempo shifting & 50 Million+ downloads with 4.7 rating at playstore \\ \hline
        Garage Band & iOS & Tempo and Pitch Adjustments & 98k+ reviews with 4.1 rating at Appstore   \\ \hline
       
        \multicolumn{4}{p{15cm}}{\textit{Data is based on Google Play Store and Apple App Store ratings and download counts as of Nov. 2024.}} 
    \end{tabular}
    \vspace{-20pt}
    \label{tab:my_label}
\end{table}
In our first step of the research, we investigate popular audio editing features that are widely available and easily accessible to users. o accomplish this, we analyzed audio editing applications available on both the Google Play Store and Apple App Store. We found that pitch and tempo adjustments are among the most commonly available and widely used features in popular audio editing applications. Using this observation, we next explore whether the adjustment of pitch and tempo provides a privacy protection against sensitive information inference attacks from audio. We hypothesize that by integrating their manipulation into the design of privacy preserving mechanism, we can achieve the dual goals of usability and privacy preservation. Since users are already familiar with these features, they are more likely to adopt and use them effectively as a privacy-preserving mechanism. To determine whether pitch and tempo manipulation can genuinely provide privacy, we designed and conducted a series of experiments. 

\begin{figure}[!h]
    \centering
    \includegraphics[width=1\linewidth]{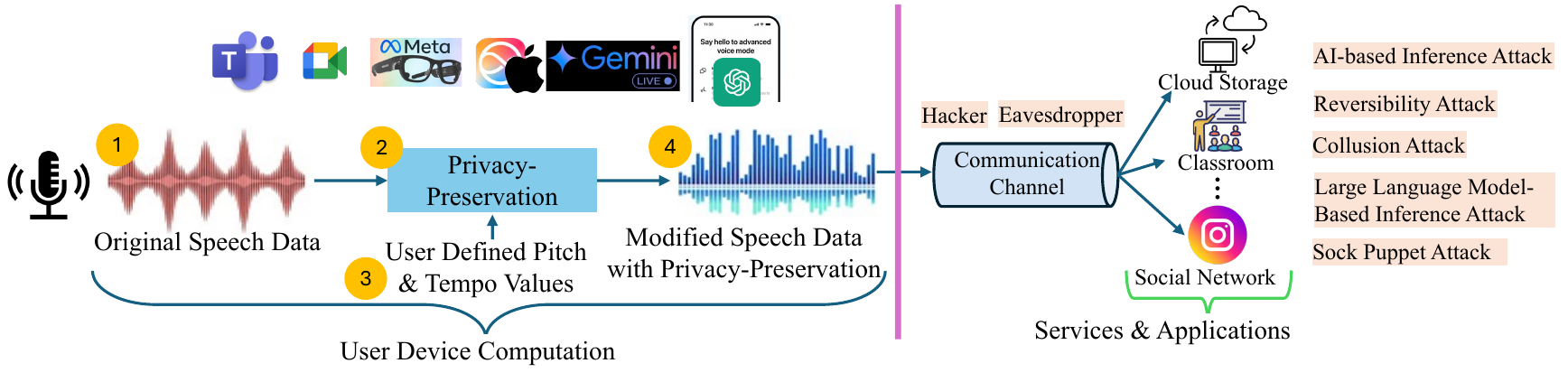} \vspace{-20pt}
    \caption{The Privacy-Preserving mechanism integrated system model. }
    \label{fig:model}
\end{figure}
Our envisioned privacy-preserving system model is depicted in Fig \ref{fig:model}, illustrating the flow from original speech data to modified data and the associated risks. Initially, the original speech data such as that from video conferencing platforms like Microsoft Teams and Google Meet or recordings captured through devices like Meta Ray-Ban, Apple Siri, or Gemini/ChatGPT undergoes user defined modifications. These modifications involve altering pitch and tempo values to different levels using practical applications as shown in Fig \ref{fig:enter-label}. Changing pitch and tempo in speech directly affects the emotion conveyed in the audio file. For example, if an audio file contains a recording in a friendly or happy tone, altering the pitch and tempo can convert it to another emotion, such as anger or disgust. The modified speech data is then transmitted through various communication channels, including user devices, computational platforms, cloud storage, and services or applications such as classrooms and social networks. Throughout this process, the framework addresses several potential privacy threats, including interception by hackers or eavesdroppers, AI-based inference attacks to extract sensitive information, reversibility attacks that attempt to reconstruct the original data, collusion attacks involving multiple entities, and inference attacks leveraging large language models. Additionally, there is a risk of sock puppet attacks, where malicious entities use fake identities to exploit the system.
\begin{figure}[h!]
    \vspace{-20pt}
    \centering
    \rotatebox{90}{%
        \includegraphics[width=0.15\linewidth]{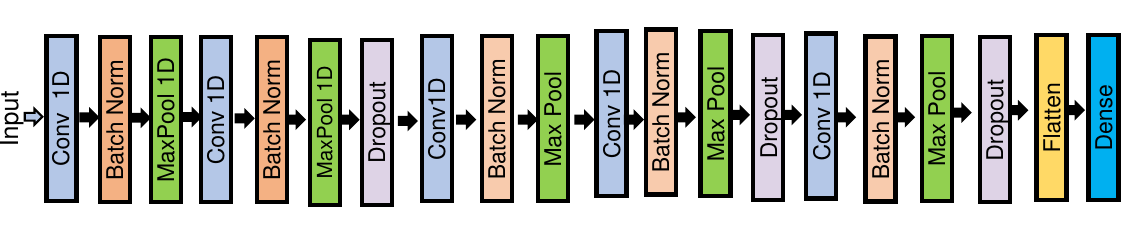}
    }
    \vspace{-15pt}
    \caption{1D-CNN Model Architecture}
    \label{fig:cnn-architecture-horizontal}

\end{figure}

\subsection{Trained Threat Models} 
As our potential attacker possesses the capability of using deep learning model including immensely powerful LLM models for the aforementioned inference attack, for attack simulation first we designed a 1-D CNN model from scratch and trained it using the RAVDESS, CREMA-D and  three prominent dataset for emotion recognition. The reason for using 1-D CNN is that not only it is lightweight which increases the research scope in that direction but also in recent researches with audio and speech data it showed impressive performance as demonstrated in \cite{10.1145/3411495.3421355}. As expected, we achieved a significantly high 97\% of test accuracy for the model. Fig \ref{fig:cnn-architecture-horizontal} shows the architecture of our 1 dimensional convolutional neural network model.

\section{Experiment and Results}

\textbf{Setup and Datasets:} In this study, we analyzed popular audio editing applications, such as Uptempo, available on the Google Play Store and Apple App Store, to identify the common audio editing features offered by these platforms. Our objective is to assess the potential of these features as privacy-preserving tools, specifically in the context of protecting user privacy against speech emotion recognition systems. Within our threat model, we explore two primary approaches for executing attacks to evaluate the effectiveness of these features. The first method involves randomly altering pitch and
tempo values while selecting male and female actors from three datasets to examine the impact on privacy.The second method involves systematically adjusting pitch and tempo across six distinct levels to create a range of variations. For pitch, the selected values are 0.0, +4, -4, +8, and -8, while tempo is varied at 60, 80, 100, 120, and 140. These combinations yield a total of 25 unique attack scenarios. For instance, when the pitch is set to 0.0, it is paired sequentially with tempo values of 60, 80, 100, 120, and 140. Similarly, for a pitch value of -4, the same tempo values are applied. This systematic approach is consistently implemented across all pitch variations to thoroughly examine their combined effects. We use three emotion speech datasets, specifically the RAVDESS \cite{livingstone2018ryerson}, CREMA-D\cite{6849440}, and TESS\cite{SP2/E8H2MF_2020} datasets. The details of these datasets are presented in Table \ref{tab:dataset_summary}.


\begin{table}[h!]
\centering \small \setlength{\tabcolsep}{3pt}
\caption{Summary of Emotion Speech Datasets}
\begin{tabular}{|p{1.8cm}|p{1cm}|p{1.6cm}|p{1.25cm}|p{4.5cm}|p{4cm}|}
\hline
\textbf{Dataset} & \textbf{Total Files} & \textbf{Actors} & \textbf{Age Range} & \textbf{Emotions} & \textbf{File Description} \\ \hline
RAVDESS\cite{livingstone2018ryerson} & 1440 & 24 (12 male, 12 female) & - & Calm, Happy, Sad, Angry, Fearful, Surprised, Disgust & 60 trials × 24 actors; 16-bit, 48 kHz audio \\ \hline
CREMA-D\cite{6849440} & 7442 & 91 (48 male, 43 female) & 20-74 & Anger, Disgust, Fear, Happy, Neutral, Sad (Low intensity) & 12 sentences × 91 actors; multimodal data \\ \hline
TESS\cite{SP2/E8H2MF_2020} & 2800 & 2 (female) & 26, 64 & Anger, Disgust, Fear, Happiness, Pleasant Surprise, Sadness, Neutral & 200 words × 2 actresses; carrier phrase: 'Say the word...' \\ \hline
\end{tabular}%
\label{tab:dataset_summary}
\end{table}

\subsection{Randomized Approach}
To evaluate our proposed attack and defense model using the Up Tempo app, we randomly altered the pitch and tempo for different categories of emotion of the same user with different combinations that are fed to the attacker's models as inputs. To examine gender-related biases, we selected one male and one female actor from each of the aforementioned datasets and analyzed their pitch variations.
Table \ref{tab:Emotion23} shows the result for Actor 23 of REVDESS dataset on randomly selected tempo and pitch values.
The results clearly demonstrate that changed pitch and tempo is creating significant effects on attacker's inference model's performance in every case from 1-D CNNs to highly capable LLM model such as GPT-4o. Similarly, Table \ref{tab:my_label} shows the outcome for actor 16 and we observe that just like the previous actors with the randomized change in tempo and pitch values, attacker both with their 1-D CNN and GPT-4o LLM model is failing to properly extract the real emotion from the data.\\
Table \ref{tab:table7} illustrates the impact of pitch and tempo variations on a male actor (Actor ID: 1001). As observed in the previous cases, the attacker consistently infers incorrect emotional information from the speech files of this actor. Following the approach taken with the RAVDESS dataset, where we transitioned from testing a male actor (Actor 23) to a female actor (Actor 16), we conducted similar tests on a female actor from the CREMA-D dataset. Table \ref{tab:my_label11s} highlights the effects of various random combinations of pitch and tempo on the female actor (Actor ID: 1002), with the significant impact of these variations clearly evident in the results.\\
Beyond the evident failure of the proposed attacker models to correctly infer emotions, a deeper analysis of the randomized approach reveals an important insight: female actors, even when drawn from different datasets and representing entirely different individuals, exhibit notable commonalities in how their emotions are affected by pitch and tempo variations. For instance, when the original emotion is anger, a combination of high pitch and high tempo often leads to a predicted emotion of fear. Similarly, for disgust, the same combination frequently results in a shift to fear. Observations from Table \ref{tab:Emotion23} and Table \ref{tab:table7} further corroborate this trend. For example, increasing the tempo while keeping the pitch constant alters the predicted emotion from neutral to disgust. These findings suggest a strong correlation between pitch, tempo, and emotional expression in speech data, prompting a more structured and detailed analysis, which is presented in the next section.

\begin{table}[h!]
    \centering \small 
    \caption{Effect of different pitch and tempo change on Male (ID 23 RAVDESS)}
\label{tab:Emotion23}
\begin{tabular}{|p{.9cm}|p{.9cm}|p{.9cm}|p{0.9cm}|p{1.1cm}||p{0.8cm}|p{0.8cm}|p{0.9cm}|p{0.9cm}|p{1.2cm}|}
\hline
 \textbf{Pitch} &  \textbf{Tempo} & \textbf{1-D CNN} & \textbf{GPT-4o} &\textbf{Original Emotion} &  \textbf{Pitch} &  \textbf{Tempo} & \textbf{1-D CNN} & \textbf{GPT-4o} & \textbf{Original Emotion} \\
\hline
  -9.4 &   72.0 &           neutral &      neutral &   Neutral &      0.0 &    129.0 &               happy & neutral & Disgust \\
\hline
  -8.4 &   76.0 &             happy &    happy &         Fear &      0.0 &    130.0 &  fear  &   sad &        Surprise \\
\hline
  -8.4 &  100.0 &           neutral &   happy &       Neutral &      0.0 &    138.0 &               angry &       angry &        Fear \\
\hline
  -8.4 &  136.0 &          surprise &   surprise &       Neutral &      0.0 &    142.0 &             disgust &     disgust &       Neutral \\
\hline
  -8.3 &  100.0 &           disgust &    angry &      Disgust &      6.3 &    100.0 &                fear &     fear &      Disgust \\
\hline
  -7.8 &   76.0 &           neutral &   neutral   &    Disgust &      7.4 &    100.0 &            surprise &   fear     &       Fear \\
\hline
  -7.6 &   73.0 &              fear &  fear &        Surprise &      7.8 &    126.0 &                fear &     fear &      Surprise \\
\hline
  -7.5 &  100.0 &          surprise &   surprise &         Fear &      7.9 &    100.0 &            surprise &    fear &       Surprise \\
\hline
  -7.4 &  100.0 &          surprise &    fear &     Surprise &      8.1 &     61.0 &             disgust &   angry &         Neutral \\
\hline
   0.0 &   61.0 &           disgust &  sad &        Neutral &      8.1 &    126.0 &               angry &   sad &         Disgust \\
\hline
   0.0 &   68.0 &               sad &  disgust &        Disgust &      8.7 &    128.0 &                fear &    fear &           Fear \\
\hline
   0.0 &   70.0 &               sad &     sad &        Fear &     10.6 &    133.0 &            surprise &       fear &     Neutral \\
\hline
   0.0 &   72.0 &             happy &  neutral &       Surprise &     10.9 &    100.0 &             neutral &    neutral  &      Neutral \\
\hline
\end{tabular}

\end{table}

\renewcommand{\arraystretch}{.75}
\begin{table}[h!]
    \centering\small
    \caption{Effects of pitch and tempo change on Female (ID-16 RAVDESS)}
    \label{tab:my_label}
        \begin{tabular}{|p{0.9cm}|p{0.9cm}|p{0.9cm}|p{0.9cm}|p{0.98cm}||p{0.9cm}|p{0.9cm}|p{0.9cm}|p{0.9cm}|p{1.2cm}|}
\hline
\textbf{Pitch} & \textbf{Tempo} & \textbf{1-D CNN}& \textbf{GPT-4o} & \textbf{Original Emotion} & \textbf{Pitch} & \textbf{Tempo} & \textbf{1-D CNN} & \textbf{GPT-4o} & \textbf{Original Emotion} \\
\hline
   0.0 &   143 &             angry &    angry  &  Surprise &   0.0 &    66 &           neutral &  happy &       Disgust \\
\hline
   0.0 &    63 &             angry &   Disgust   &  Surprise &   0.0 &    66 &           neutral &  neutral &        Disgust \\
\hline
   9.7 &   141 &              fear &   fear &    Surprise &   0.0 &   142 &          surprise &   fear &      Disgust \\
\hline
  -8.9 &    62 &               sad &    disgust &    Surprise &   7.6 &   133 &              fear &   fear &       Disgust \\
\hline
  -8.1 &   100 &          surprise &    fear &    Surprise &  -7.5 &    70 &              fear &    fear &     Disgust \\
\hline
  10.3 &   100 &              fear &    fear &     Surprise &  -5.7 &   100 &           disgust &   angry &       Disgust \\
\hline
   0.0 &    66 &              fear &      fear &        Sad &   8.9 &   100 &             angry &    angry &     Disgust \\
\hline
   8.0 &   100 &               sad &     sad &         Sad &   9.0 &   133 &              fear &      fear &      Anger \\
\hline
  -7.9 &   100 &           disgust &       disgust &       Sad &  -8.5 &    70 &           disgust &    sad &        Anger \\
\hline
   8.3 &   130 &             angry &    angry &          Sad &   0.0 &   132 &             angry &    angry &       Anger \\
\hline
  -9.1 &    68 &              fear &    surprise &          Sad &   0.0 &    62 &              fear &     fear &       Anger \\
\hline
   0.0 &   127 &             angry &        angry &      Sad &   7.6 &   100 &               sad &    disgust &       Anger \\
\hline
   7.2 &   100 &               sad &    disgust &    Neutral &  -7.0 &   100 &             angry &   sad &         Anger \\
\hline
\end{tabular}
    
\end{table}

\begin{table}[h!]
    \centering\small
    \caption{Impact of Pitch and Tempo Variations on Male Actor (ID:1001 from CREMA-D)}
    \label{tab:table7}
\begin{tabular}{|p{.9cm}|p{.9cm}|p{0.9cm}|p{0.9cm}|p{1.2cm}||p{0.9cm}|p{0.9cm}|p{0.9cm}|p{0.9cm}|p{1.2cm}|}
\hline
 \textbf{Pitch} &  \textbf{Tempo} & \textbf{1-D CNN} & \textbf{GPT-4o}& \textbf{Original Emotion} &  \textbf{Pitch} &  \textbf{Tempo} & \textbf{1-D CNN} & \textbf{GPT-4o} & \textbf{Original Emotion} \\
 \hline
   0.0 &     64 &             happy &     neutral &        Fear &    0.0 &     65 &           disgust &    sad &      Disgust \\
   \hline
   0.0 &    133 &           disgust &     sad &        Fear &    0.0 &    132 &           disgust &   angry &       Disgust \\
   \hline
  -8.5 &     68 &           disgust &     disgust     &   Fear &   -8.8 &     68 &           neutral &   happy &       Disgust \\
  \hline
   8.2 &    133 &           disgust &    disgust &         Fear &    8.7 &    134 &             happy &  happy &       Disgust \\
   \hline
  -8.5 &    100 &              fear &       fear &      Fear &   -7.6 &    100 &             happy &   neutral &       Disgust \\
  \hline
   8.3 &    100 &              fear &     sad &       Fear &    7.7 &    100 &           disgust &    angry &      Disgust \\
   \hline
   0.0 &     64 &             happy &       neutral   &   Sad &    0.0 &     67 &              fear &   fear &         Happy \\
   \hline
   0.0 &    136 &           disgust &      disgust &        Sad &    0.0 &    131 &           disgust &   angry &         Happy \\
   \hline
  -7.8 &     70 &             happy &     neutral &         Sad &   -8.2 &     70 &             happy &     neutral &       Happy \\
  \hline
   8.5 &    132 &              fear &    sad &          Sad &    8.3 &    131 &              fear &     fear &       Happy \\
   \hline
  -8.2 &    100 &              fear &    fear &          Sad &   -8.6 &    100 &             happy &    happy &        Happy \\
  \hline
   8.1 &    100 &           disgust &    disgust &          Sad &    7.8 &    100 &           disgust &   angry   &      Happy \\
   \hline
   0.0 &     68 &             angry &   angry &         Anger &    0.0 &     67 &              fear &   fear &       Neutral \\
   \hline
   0.0 &    136 &              fear &     fear &      Anger &    0.0 &    131 &           disgust &   angry    &   Neutral \\
   \hline
  -9.1 &     66 &             angry &     disgust &       Anger &   -8.2 &     70 &             happy &  neutral &      Neutral \\
  \hline
   7.4 &    136 &              fear &     fear &       Anger &    8.3 &    131 &              fear &    fear   &s   Neutral \\
   \hline
  -7.6 &    100 &           neutral &     happy &       Anger &   -8.6 &    100 &             happy &  happy    &    Neutral \\
  \hline
   7.8 &    100 &             angry &    angry &       Anger &    7.8 &    100 &           disgust &  angry   &     Neutral \\
   \hline
\end{tabular}
\end{table}

\begin{table}[h!]
    \centering\small
    \caption{Impact of Pitch and Tempo Variations on Female (ID-1002 CREMA-D).}
    \label{tab:my_label11s}
\begin{tabular}
{|p{.7cm}|p{.9cm}|p{0.9cm}|p{0.98cm}|p{1.2cm}||p{0.9cm}|p{0.9cm}|p{0.9cm}|p{0.9cm}|p{1.2cm}|}
\hline
\textbf{Pitch} & \textbf{Tempo} & \textbf{1-D CNN} & \textbf{GPT-4o} & \textbf{Original Emotion} & \textbf{Pitch} & \textbf{Tempo} & \textbf{1-D CNN} & \textbf{GPT-4o} & \textbf{Original Emotion} \\
\hline
   0.0 &     63 &             angry &  angry    &    Disgust &    0.0 &     66 &           disgust &  Neutral &        Neutral  \\
\hline
   0.0 &    132 &             angry &   angry &      Disgust &    0.0 &    133 &             happy &    happy &      Neutral \\
\hline
  -9.1 &     68 &             angry &    sad &      Disgust &   -8.5 &     68 &             angry &   disgust &       Neutral \\
\hline
   8.8 &    130 &              fear &   fear &      Disgust &    8.4 &    131 &             happy &    happy &      Neutral \\
\hline
  -7.0 &    100 &             angry &   angry  &      Disgust &   -8.2 &    100 &           neutral &   happy &       Neutral \\
\hline
   7.9 &    100 &              fear &   fear &      Disgust &    8.5 &    100 &             happy &    happy &      Neutral \\
\hline
   0.0 &     68 &               sad &     sad &        Fear &    0.0 &    137 &              fear &      disgust &        Sad \\
\hline
   0.0 &    131 &              fear &     fear &        Fear &   -8.7 &     70 &           disgust &    angry &          Sad \\
\hline
  -8.1 &     70 &           disgust &   angry &          Fear &    9.0 &    134 &          surprise &     disgust &         Sad \\
\hline
   7.0 &    132 &             angry &    disgust &         Fear &   -7.7 &    100 &           neutral &    happy &          Sad \\
\hline
  -8.2 &    100 &           neutral &    sad &        Fear &    8.9 &    100 &               sad &   sad &           Sad \\
\hline
   8.0 &    100 &               sad &   disgust    &   Fear &    0.0 &     68 &           disgust &      angry &        Sad \\
\hline
   0.0 &     66 &              fear &    fear &        Happy &    0.0 &     63 &             angry &    disgust &        Anger \\
\hline
   0.0 &    132 &             happy &     happy &       Happy &    0.0 &    132 &             angry &     angry &       Anger \\
\hline
  -8.0 &     64 &              fear &     fear &       Happy &   -9.1 &     68 &             angry &   disgust &         Anger \\
\hline
   8.6 &    131 &             happy &     neutral &       Happy &    8.8 &    130 &              fear &     surprise &       Anger \\
\hline
  -7.8 &    100 &             angry &     disgust   &    Happy &   -7.0 &    100 &             angry &     angry &       Anger \\
\hline
   8.5 &    100 &             happy &     neutral &       Happy &    7.9 &    100 &              fear &    fear &        Anger \\
\hline
\end{tabular}
 
\end{table}

\begin{table}[h!]
    \centering\small
    \caption{Impact of Pitch and Tempo Variations on Female Actor OAF of Tess Dataset}
    \label{tab:my_label11s}
\begin{tabular}
{|p{.7cm}|p{.9cm}|p{0.9cm}|p{0.98cm}|p{1.2cm}||p{0.9cm}|p{0.9cm}|p{0.9cm}|p{0.9cm}|p{1.2cm}|}
\hline
\textbf{Pitch} & \textbf{Tempo} & \textbf{1-D CNN} & \textbf{GPT-4o} & \textbf{Original Emotion} & \textbf{Pitch} & \textbf{Tempo} & \textbf{1-D CNN} & \textbf{GPT-4o} & \textbf{Original Emotion} \\
\hline
   0.0 &     84 &             angry &  sad    &    Disgust &    0.0 &     70 &           disgust &  Neutral &        Neutral  \\
\hline
   0.0 &    130 &             angry &   sad &      Disgust &    0.0 &    136 &             happy &    happy &      Neutral \\
\hline
  -9.1 &     68 &             angry &    sad &      Disgust &   -7.5 &     68 &             angry &   angry &       Neutral \\
\hline
   8.5 &    133 &              fear &   fear &      Disgust &    8.4 &    131 &             happy &    neutral &      Neutral \\
\hline
  -6.5 &    100 &             angry &   angry  &      Disgust &   -8.2 &    100 &           neutral &   happy &       Neutral \\
\hline
   7.9 &    100 &              fear &   fear &      Disgust &    8.2 &    100 &             happy &    happy &      Neutral \\
\hline
   0.0 &     68 &               sad &     sad &        Fear &    0.0 &    137 &              fear &      fear &        Sad \\
\hline
   0.0 &    131 &              sad &     disgust &        Fear &   -8.7 &     70 &           sad &    angry &          Sad \\
\hline
  -6.9 &     72 &           surprise &   sad &          Fear &    8.00 &    134 &          disgust &   neutral  &         Sad \\
\hline
   7.0 &    132 &             angry &    disgust &         Fear &   -7.7 &    100 &           neutral &    happy &          Sad \\
\hline
  -8.2 &    100 &           neutral &    sad &        Fear &    8.00&    100 &               sad &   sad &           Sad \\
\hline
   8.0 &    100 &               sad &   disgust    &   Fear &    0.0 &     68 &           disgust &      angry &        Sad \\
\hline
   0.0 &     65 &              fear &    fear &        Happy &    0.0 &     62 &             angry &    disgust &        Anger \\
\hline
   0.0 &    132 &             happy &     happy &       Happy &    0.0 &    132 &             angry &     angry &       Anger \\
\hline
  -8.0 &     64 &              fear &     fear &       Happy &   -8.3 &     65 &             angry &  angry  &         Anger \\
\hline
   8.6 &    131 &             happy &     happy &       Happy &    7.5 &    130 &              fear &     surprise &       Anger \\
\hline
  -7.5 &    123 &             angry &     disgust   &    Happy &   -8.0 &    100 &             angry &     anger &       Anger \\
\hline
   8.5 &    100 &             happy &     happy &       Happy &    7 &    100 &              fear &    fear &        Anger \\
\hline
\end{tabular}
 
\end{table}
\subsection{Structured Approach}
While the goal of randomized approach was to see whether variations in pitch and combo can make the attacker extract incorrect emotions, the goals of this structured section is to provide a brief analysis to find out how this emotion change happens. Unlike the randomized approach, tempo and pitch values are not selected randomly this, but with regular interval. As previously mentioned, we have used pitch value of -8,-4,0,4,8 and tempo values of 60,80,100,120,140. For each of the emotion of category we made possible combinations with the designated pitch and tempo values and figured out the attacker model's response. Here, we will using previously discussed female actor (ID-16 RAVDESS) speech audio with Neutral emotion  as a test case. Table \ref{tab:neutral} demonstrates the the pitch and tempo variations on actor 16's speech with neutral emotion. And it represent how different combinations of pitch and tempo influence emotion classification by two models: a 1-D CNN (Convolutional Neural Network) and GPT-4o, a variant of GPT-4 optimized for this task. Both models occasionally agree on the emotional label (e.g., at 60 BPM and -8 pitch, both predict ``sad"). Lower tempos (e.g., 60 BPM) and negative pitch shifts (-8, -4) often correlate with ``sad" or ``disgust." Higher tempos (e.g., 120, 140 BPM) and positive pitch shifts (4, 8) show varied predictions like ``happy," ``fear," or ``surprise." Neutral tempo (0 BPM) shows variability in emotion predictions, with both ``neutral" and other emotions like ``happy" appearing. Table \ref{tab:sad_combined} represents for both models predict ``Sad" emotion reliably at a moderate pitch (0 or 8) and moderate tempo (60 or 80 BPM). There is significant difference between 1-D CNN and GPT-4o, especially at higher tempos (100–140 BPM), where GPT-4o often predicts more varied emotions like ``happy" or ``surprise." and 1-D CNN maintains some consistency in identifying ``neutral" or ``sad" in high-tempo scenarios.

\begin{table}[h!]
\centering \small
\caption{Effects of Pitch and Tempo variations on \textbf{Neutral} Emotion.}
\label{tab:neutral}
\begin{tabular}{|p{0.9cm}|p{0.9cm}|p{0.9cm}|p{0.9cm}||p{0.9cm}|p{0.9cm}|p{0.9cm}|p{0.9cm}|}
\hline
\textbf{Tempo} & \textbf{Pitch} & \textbf{1-D CNN} &\textbf{GPT-4o} & \textbf{Tempo} & \textbf{Pitch} & \textbf{1-D CNN} & \textbf{GPT-4o} \\ \hline
60             & -8             & sad     & sad         & 140            & 8              & fear   & fear          \\ \hline
60             & -4             & sad    & sad         & 140            & 8              & fear   & surprise          \\ \hline
60             & 8              & sad  &   sad         & 140            & 4              & fear    & fear         \\ \hline
60             & 4              & sad   & neutral           & 140            & 0              & fear     & fear        \\ \hline
60             & 0              & sad   &    sad        & 120            & -8             & disgust  & anger        \\ \hline
80             & -8             & disgust   & anger       & 120            & -4             & angry   & anger         \\ \hline
80             & -4             & disgust    & disgust      & 120            & -4             & angry   & anger        \\ \hline
80             & 8              & disgust   & anger       & 120            & 4              & angry   & sad         \\ \hline
80             & 4              & happy   & happy         & 120            & 0              & angry    & sad        \\ \hline
80             & 0              & disgust    & anger      & 0              & -8             & neutral & happy         \\ \hline
140            & -8             & sad    & sad          & 0              & -4             & neutral   & happy       \\ \hline
140            & -8             & sad   & neutral           & 0              & 8              & neutral  & neutral         \\ \hline
140            & -4             & fear    & surprise         & 0              & 4              & neutral & neutral         \\ \hline
\end{tabular}

\end{table}

\begin{table}[h!]
\centering \small
    \caption{Effects of Pitch and Tempo on \textbf{Sad} Emotion}
\label{tab:sad_combined}
\begin{tabular}{|p{0.9cm}|p{0.9cm}|p{0.9cm}|p{0.9cm}||p{0.9cm}|p{0.9cm}|p{0.9cm}|p{0.9cm}|p{0.9cm}|}
\hline
\textbf{Tempo} & \textbf{Pitch} & \textbf{1-D CNN} &\textbf{GPT-4o} & \textbf{Tempo} & \textbf{Pitch} & \textbf{1-D CNN} & \textbf{GPT-4o} \\ \hline
60 & -8 & disgust & anger & 140 & 8 & fear & fear \\ \hline
60 & -4 & sad & sad& 140 & 4 & fear & surprise\\ \hline
60 & 8 & sad & sad & 140 & 0 & fear & surprise\\ \hline
60 & 4 & sad & disgust& 120 & -8 & neutral & neutral \\ \hline
60 & 0 & sad & disgust & 120 & -4 & neutral & neutral \\ \hline
80 & -8 & disgust & anger & 120 & 8 & surprise & surprise \\ \hline
80 & -4 & disgust & anger & 120 & 4 & surprise & anger\\ \hline
80 & 8 & sad & sad & 120 & 0 & surprise & surprise\\ \hline
80 & 4 & disgust & anger & 100 & -8 & neutral & happy \\ \hline
80 & 0 & disgust & anger & 100 & -4 & neutral & sad\\ \hline
140 & -8 & neutral & happy & 100 & 8 & sad & sad\\ \hline
140 & -4 & happy & happy & 100 & 4 & neutral & neutral \\ \hline
\end{tabular}
\end{table}

\section{Conclusion and Future Work}

In this paper, we investigate the impact of modifying the pitch and tempo of speech audio data using user-accessible apps from the Google Play Store and iOS App Store. Our findings indicate that the evaluated attack models consistently failed to accurately infer emotions from the modified data, effectively providing privacy protection for sensitive user speech data. This demonstrates a practical, user-centric appraoch to protect privacy of sensitive data from speech audio in the face of adversaries with capabilities of state-of-the-art inference models, including multimodal large language models. The novelty of this work lies in its practical contribution to bridging the gap between usability and privacy, offering users a promising method to protect their speech data through familiar applications. 

For future work, we plan to extend the scope of this research in several important directions. We plan to simulate more advanced attackers leveraging other LLMs, such as Flan-T5, Gemma-2B, and LLama, fine-tuned specifically for speech emotion recognition (SER) tasks. We also plan to integrate recent SER advancements, including universal audio representation models like Emotion2Vec2 \cite{ma2023emotion2vecselfsupervisedpretrainingspeech} and Wave2Vec2 \cite{baevski2020wav2vec20frameworkselfsupervised}, into our attack framework, to facilitate the development of more complex and advanced attack scenarios to get better insights into the robustness and privacy implications of speech data modifications. We will also investigate the reversibility of pitch and tempo modifications using signal-processing techniques and how to mitigate that. Furthermore, we plan to assess privacy protection for additional attributes (e.g. identify or gender) beyond emotion inference.

\bibliographystyle{unsrt}  
\bibliography{references}

\end{document}